# The Geobiology of Weathering: a 13th Hypothesis*

David W. Schwartzman
Department of Biology, Howard Univ., Washington DC 20059, USA
(dschwartzman@gmail.com)

## Abstract

The magnitude of the biotic enhancement of weathering (BEW) has profound implications for the long-term carbon cycle. The BEW ratio is defined as how much faster the silicate weathering carbon sink is under biotic conditions than under abiotic conditions at the same atmospheric $pCO_2$ level and surface temperature. Thus, a $13^{th}$ hypothesis should be considered in addition to the 12 outlined by Brantley et al. (2011) regarding the geobiology of weathering: The BEW factor and its evolution over geological time can be inferred from "meta-analysis" of empirical and theoretical weathering studies. Estimates of the global magnitude of the BEW are presented, drawing from lab, field, watershed data and models of the long-term carbon cycle, with values ranging from one to two orders of magnitude.

## Introduction

Twelve hypotheses were advanced to promote critical zone research relevant to the geobiology of weathering (Brantley et al., 2011).

A 13th hypothesis is proposed:

The biotic enhancement of weathering (BEW) factor and its evolution over geological time can be inferred from "meta-analysis" of empirical and theoretical weathering studies. Estimates of the global magnitude of the BEW can be inferred from lab, field, watershed data and models of the long-term carbon cycle. The BEW ratio is defined as how much faster the silicate weathering carbon sink is under biotic conditions than under abiotic conditions at the same atmospheric $pCO_2$ level and surface temperature (Schwartzman and Volk (1989).

The likely contributors to the present BEW with forest and grassland ecosystems/soils as the main sites include (Schwartzman, 2008):

1. Soil stabilization (likely the most important, and likely more than a tenfold contribution to reactive mineral surface);

\* *(Note: this is an updated and revised version of what I presented as the Keynote for Biotic Enhancement of Weathering Session, Goldschmidt 2013, see Schwartzman and Brantley, 2013)*

2. pCO$_2$ elevation in soil from root/microbial respiration and decay (Lovelock and Watson's (1982) original BEW);
3. Organic acids/chelators in soil, the multifold interactions in the rhizosphere/mycorrhizae including biogeophysical/biochemical weathering by soil fungi driving the breakup and digestion of CaMg silicate mineral particles (e.g., see Bray et al., 2013).
4. Evapotranspiration contribution to maintaining soil water and runoff (plant biological pump, Lucas, 2001);
5. Diffusion of oxygen into soil contributing to oxidation of CaMgFe silicates and production of sulfuric and nitric acids;
6. Water retention by soil organics;
7. Turnover of soil by ants, earthworms, mixing organic and mineral particles (e.g., Darwin's pioneering research); and
8. Biotic sink effect for mineral nutrients (assuming steady-state soil/biomass (with net flux of Ca, bicarbonate out), or increasing plant biomass storing Ca to be released with organic decay)

Recent research points to the role of ants as strong agents of chemical weathering of silicates (Dorn, 2014).

There are also biotic effects that may slow down weathering:

1. Accumulation of thick, depleted soils in low slope terrains acting as a barrier to water flow to fresh bedrock, although the role of vegetative cover is unclear. Forest canopy slows down raindrop erosion, but there is evidence of significant bedrock-derived chemical weathering in lowland humid tropical weathering, plausibly enhanced by the biota (Porder et al., 2006) compared to abiotic exposed bedrock.
2. Macropores created by earthworms, ants and other animals in soils that allow water flow bypassing saturated pore space.
Lichen coverage of bedrock in most cases promotes chemical denudation, so it is not an example of biotic retardation of weathering (BRW).
But even with these retarding effects, BEW dominates globally.

An abiotic Earth surface has no soil, little regolith, so the land surface potential reacting with water/carbon dioxide is closer to two-dimensional, with a lower surface roughness.

Biological weathering and mycorrhizal evolution include the following mechanisms (Taylor et al., 2009):

1: Acidification due to hydrogen ion and organic exudates
2: Acidification due to respiration and elevated PCO$_2$(g)
3: Litter decomposition and carbon transfer to heterotrophs
4 and 5: Evapotranspiration and stabilization of soil

Here is a corollary hypotheses relating weathering, erosion and biology from Brantley et al. (2011):

Biology-driven weathering dominates the weathering sink of carbon in the long-term carbon cycle. In eroding landscapes, weathering front advance at depth is coupled to surface denudation via biotic processes (hypothesis 4).  Note that coupling of rapid erosion and weathering (high slope) feeding sediments to river floodplains with apparent comparable carbon dioxide consumption rates from silicate weathering (Bouchez et al., 2013; Moquet et al., 2011, 2014).

**Table 1. The evolution of biota and the biosphere: the increase of BEW over geologic time**

**Hadean/early Archean**
    Procaryotic soil crusts, invasion into critical zone (methanogens, anoxygenic phototrophs, early heterotrophs); soil stabilization and microbial dissolution

**Archean**

2.8 Ga   Cyanobacteria (boost in productivity on land, oxygenic photosynthesis)
2.4 Ga   Rise of atmospheric oxygen;  oxidation in critical zone weathering (e.g., promoting spheroidal weathering, microfracturing, Navarre-Sitchler et al., 2013)

**Late Archean, Proterozoic**
Ice wedging, promoting physical weathering, erosion (included as BEW process because of BEW-promoted climate cooling reached the water freezing limit on mountains); eucaryote/prokaryote symbioses on land (increase in biotic productivity)

**Late Proterozoic, Phanerozoic**
Fungi (lichens), Bryophytes, Vascular Plants, Forests
Rhizosphere and mychorrhizal evolution, first arbuscular mycorrhizal fungi, followed by ectomycorrhizal fungi in Cenozoic (Taylor et al., 2009)
Further growth in biotic productivity on land drives higher critical zone $pCO_2$ levels and amplifies the plant biological pump (Lucas, 2001)

**Lichen Weathering**

Lichens are now the dominant vegetation of 8% of the land surface (Larson, 1987), with likely even greater importance in the late Proterozoic, prior to the emergence of plants.

**Field/Experimental (Mesocosm) Estimates**

**Table 2. Studies measuring lichen weathering flux**

Lichens (Chemical) weathering enhancement
Hubbard Brook, New Hampshire, U.S.: Metasedimentary quartz and mica schist
(Aghamiri and Schwartzman, 2002; see Fig. 1)
Mini-watersheds in field
16.0 (Mg)
4.4 (Si)
0.09 (Fe)
Wanaque, New Jersey, U.S.: Hornblende Granite
(Zambell et al., 2012)
Mini-watersheds in field
2.5 (Mg)
3.5 (Ca)
1.9 (Si)
0.5 (Fe)

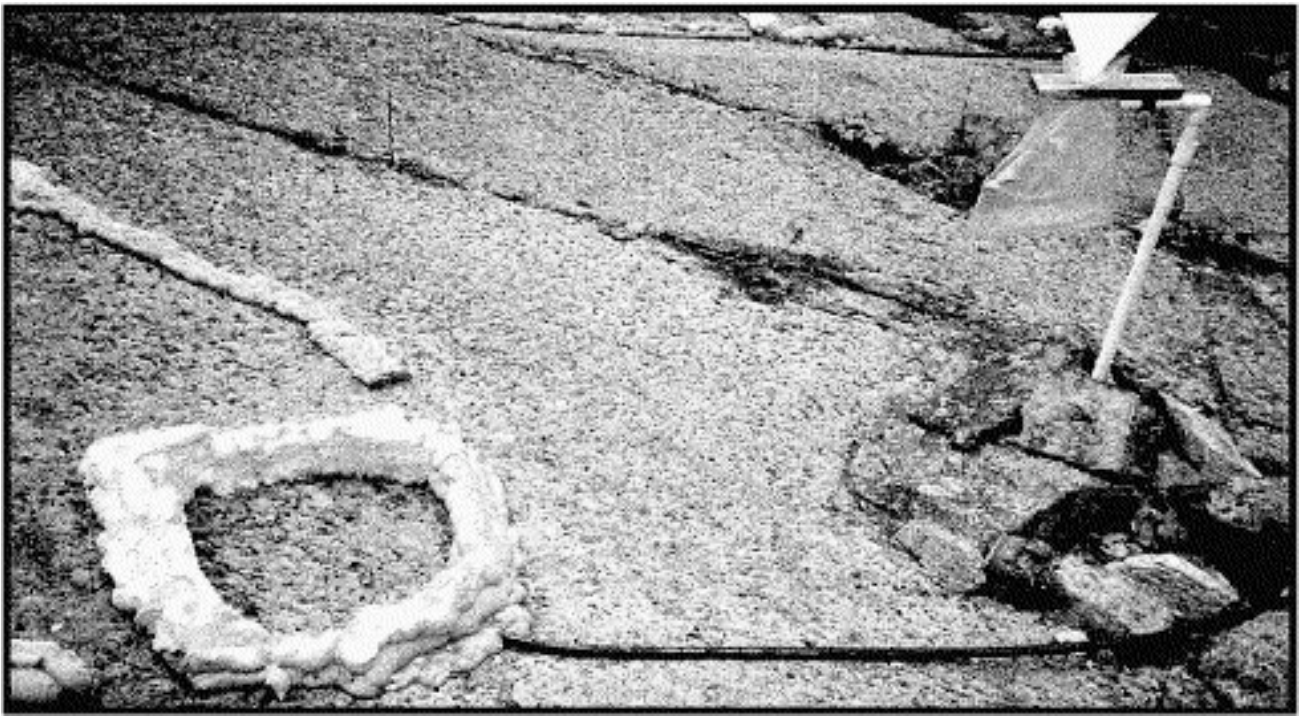

**Figure 1. Lichen mini watershed <1 m$^2$ and precipitation collection**
Hubbard Brook, New Hampshire (Aghamiri and Schwartzman, 2002)

**Table 3. Studies on lichen-induced weathering rinds**
                (Chemical) weathering enhancement factor

Mauna Loa and Kilauea, Hawaii: Basalt flows    15–71
(Jackson and Keller, 1970)

Hualalai Volcano, Hawaii: Basalt              2–18

(Brady et al., 1999)

Lanzarote, Canary Islands Basalt                16
(Stretch and Viles, 2002)
(Source: Zambell et al., 2012)

**Table 4. Mosses: (Microcosm experiment)**
            **(Chemical) weathering enhancement factor**

|          | Ca         | Mg         |
|----------|------------|------------|
| Granite  | 1.4 ± 0.2  | 1.5 ± 0.2  |
| Andesite | 3.6 ± 0.9  | 5.4 ± 0.9  |

(Lenton et al., 2012)

**Table 5. Vascular plants chemical weathering**
                **(Chemical) weathering enhancement factor**

Skorradulur, Iceland:
Basalt  (Moulton and Berner, 1998)
             3–5 (Mg)
             2–3 (Ca)
             2–3 (Si)
Hubbard Brook, NH:
Granitic outwash (manufactured soil)     (Bormann et al., 1998)
             18 (Mg)
             10 (Ca)

*Note that recent research points to a one-order BEW of vascular plants over "bare" (at least partially lichen and bryophyte-covered) bedrock  (Amundson et al., 2015; Hahm et al., 2014).*

**Inferences from Global Biogeochemical Models**

While mesocosm/field studies capture local, likely short-term BEW, global models have the potential to capture long term global BEW.

**Table 6. Global modeling of BEW**

**Phanerozoic**

1) Berner lab, culminating in GEOCARB III (2001), GEOCARBSULF (2006)
Devonian large vascular land plants weathering factor    4 x

(But more recent papers suggest this is a minimum because atmospheric $pCO_2$ levels at about 400 mya may have been higher than Berner assumed in 2001 for his weathering factor estimate, e.g., Royer et al., 2004)

2) Beerling lab
Cenozoic Plant Weathering Factor / Berner's plant factor   2-4 x
(From Figure 8. Taylor et al., 2012)

3) Kleidon lab  (Arens and Kleidon, 2011)

Removed soil $pCO_2$ elevation factor from global land Ca flux:
$$\text{maximum effect 5 x}$$
(Effect limited to regions with active erosion)

**Precambrian**

1) Von Bloh lab
Phanerozoic BEW 5.4 x higher than Precambrian
(But suggests progressive growth of BEW in Phanerozoic)
(Von Bloh et al., 2008)

2) Schwartzman  (1999 2002)
Model calculations assuming a hot Archean climate (2.6 to 3.5 Ga):
Present / Archean BEW: 25 to 82

**Modeling the variation of BEW over Geologic Time**

See my updated case for a hot Archean climate with temperatures on the order of 50-70 deg C (Schwartzman, 2015). What follows assumes this temperature scenario.

$B_R$ is defined as equal to $B_{Now} / B_t$, the ratio of the present BEW to that in the past.
As modeled in Schwartzman (1999 2002); Schwartzman and Volk (1989); Schwartzman and Volk (1991),

$$B_R = (A/A_o)(V_o/V) [(P_t/ P_o)^a (e^{\beta(T_t-T_o)}) ( e^{\gamma(T_t-T_o)} )]$$

Where $(V/V_o)$ is the volcanic outgassing flux of $CO_2$ ratio, $(A/A_o)$ is the land area available for weathering ratio, $(P_t/ P_o)$ is the ratio of atmospheric $pCO_2$, $T_t$ and $T_o$ are surface temperatures all at time t relative to the present.

Considering the sensitivity of computed $B_R$ values:
At time t, the higher the assumed $V/V_o$ ratio, the lower the computed $B_R$ value (greater outgassing drives higher temperatures), and the higher the assumed A/A₀ ratio the higher the computed $B_R$ value (greater land area drives lower temperatures.

Table 7. Model Results from Schwartzman (1999, 2002)

DID SURFACE TEMPERATURES CONSTRAIN MICROBIAL EVOLUTION? 150

TABLE 8-4.
Computed Model Results

| $T$ (°C) | Atm. p$CO_2$ (bars) | Time B.P (b.y.) | Required $B_R$ | | |
|---|---|---|---|---|---|
| | | | a | b | c |
| 70 | 2.56 | 3.5 | 658 | 82 | 25 |
| 60 | 1.22 | 2.6 | 301 | 82 | 43 |
| 50 | .31 | 1.5 | 103 | 53 | 38 |

TABLE 8-3.
Parameters for Assumed Variation of Land Area (A) and Carbon Dioxide Outgassing Rate (V) as a Function of Age (t)

| Model | $(V/V_o)(A_o/A)$ at $t$ = 3.8 Ga | $cV_o$ | ω |
|---|---|---|---|
| a (constant) | 1 × 1 = 1 | – | 0 |
| b (preferred) | 3 × 4 = 12 | 0.375 | 0.289 |
| c (upper limit) | 8 × 10 = 80 | 0.129 | 0.547 |

**Note that model c in Table 7 is now supported with respect to the ratio of land area by Flament et al. (2013 ) and the ratio of volcanic outgassing rate by Jellinek and Jackson (2015); also see references in Schwartzman (2015).**

**Revisiting the modeling in light of research regarding the variation of likely land area available for weathering (A/A$_o$) back in the Precambrian (Flament et al., 2013)**

**Table 8. Model Calculations**

| Time (Ga) | b $B_R$ | c | (A/A$_o$)* | $B_R$ b** | c** |
|---|---|---|---|---|---|
| 3.5 | 82 | 25 | 0-0.2 | 0-48 | 0-20 |
| 2.6 | 82 | 43 | 0.18-0.28 | 25-40 | 13-20 |
| 1.5 | 53 | 38 | 0.39-0.45 | 26-30 | 18-20 |

\* Derived from Flament et al. (2013)
\*\* Assuming the original V$_o$/V ratios.

Assuming Flament et al.'s (2013) constraints on emerged continental area ("A"), taking into account the weatherability factor, higher weatherability is expected in Archean, because of more mafic continental crust and greater area of subaerial oceanic ridges (and oceanic islands of basaltic/komatiitic composition). This effect could be very significant, noting that basalt has a 5-10 times greater weatherability than granite (Dessert et al., 2003; Dupre et al., 2003). Higher Archean land weatherability would increase the computed $B_R$ values, since this would be equivalent to greater effective land area available for chemical weathering. With the assumptions embedded in this modeling, including hot Archean climates, estimated land areas in the Archean derived from Flament et al. (2013) give minimum values of present BEW of 20-48.

**Table 9. Tentative estimates of cumulative BEW, with progressive increase to present**

Microbial land/critical zone biota + atmospheric oxygen + ice wedging:  5 to 10 (?)

(Eucaryotic algae, lichens, bryophyte land biota: 2 to 5, largely replaced by

Vascular plant ecosystems, culminating in forests and grasslands: 10

Cumulative:  5 x 10= 50

**Some caveats**

Global models should take into account other carbon sinks in the long-term carbon cycle, i.e., burial of reduced organic carbon, sea-floor and hydrothermal weathering, in particular their variation over geologic time compared to the critical zone weathering sink. For example, a greater net reduced organic carbon burial now relative to the past would make computed $B_R$ values too high, while a greater relative role of volcanic, sea floor and impact-derived regolith weathering in the Archean would make computed $B_R$ values too low (see Schwartzman, 1999 2002), again with the assumption of a hot Archean climate scenario.

**Meta-analysis of weathering studies at scales from the lab to the watershed; can the present BEW level be inferred ?**

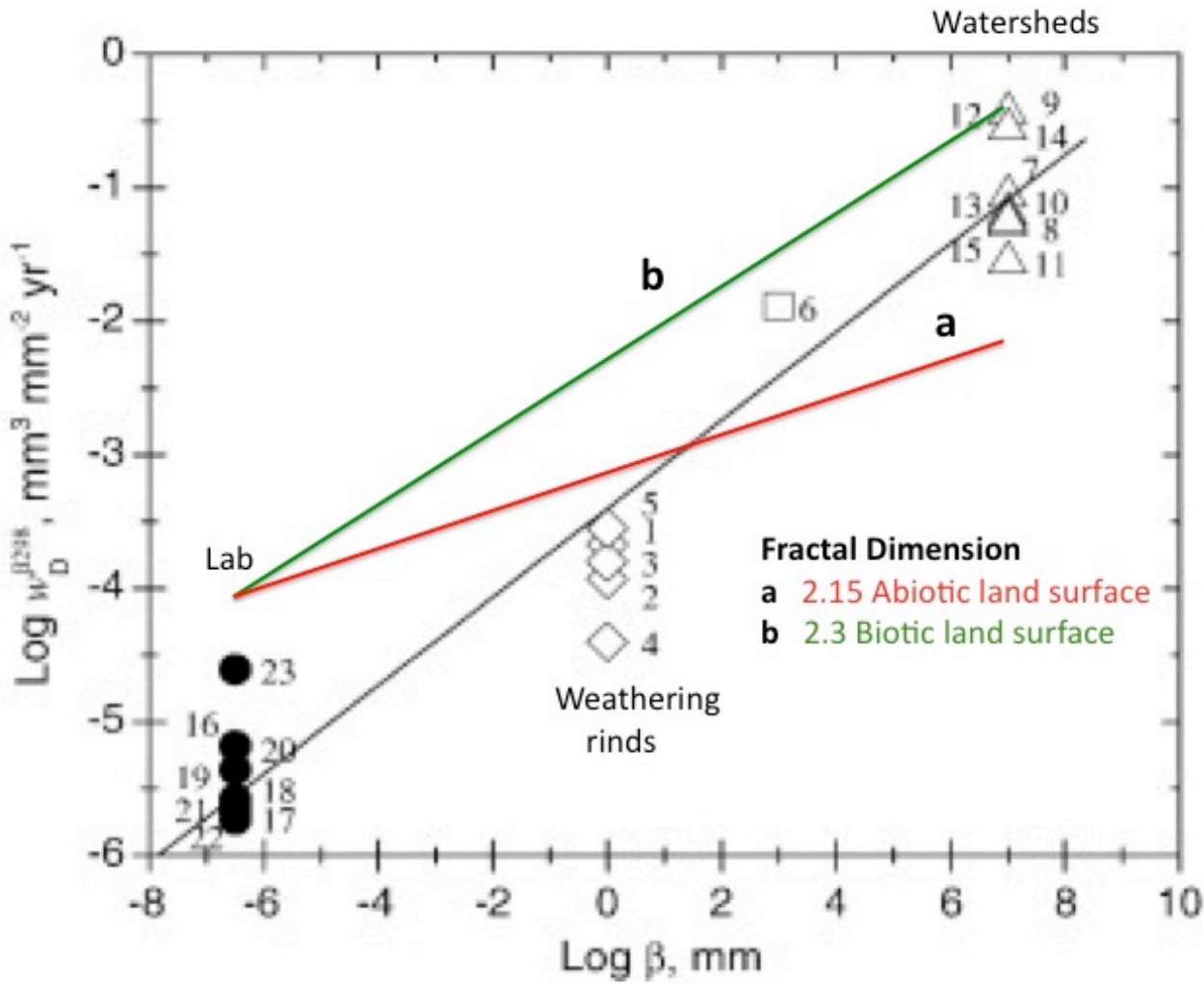

**Figure 2. Basalt Weathering (time and scale of measurement)**
Y axis: Weathering advance rate
X axis: Scale, β in mm
From Figure 3, Navarre-Sitchler and Brantley (2007), lines a and b added.

Given the evidence for a significant artifact effect on dissolution rates derived from ground powders (e.g., Eggleston et al., 1989) and for parabolically decreasing rates over year timescales in lab experiments (White and Brantley, 2003), I propose using data from a beach sand (not pulverized) experiment on forsterite (Grandstaff, 1986). I note that olivine has the highest dissolution rate of silicates making up basalt. From Grandstaff's data, I computed an estimated *geometrically-estimated* surface area and *weathering advance rate of $1 \times 10^{-4}$ mm/year* (plotted on Figure 2). I recognize that using this value is roughly 0.1 the rate derived from Rimstidt et al.'s (2012) meta-analysis of forsterite dissolution rates at a pH of 5.5 and 25 deg C, from their Fig.13 and 10. I used this computed rate as an anchor for both field and a proposed abiotic lines to derive an estimated BEW value of 41 at the watershed scale (hot humid climate, data point 9). The abiotic line was generated assuming a roughness fractal dimension of 2.15 corresponding to a sand surface without visible biota (Gao et al., 2014). Is this close to a fractal dimension of an abiotic Earth land surface at near modern $pCO_2$ and temperature levels under consideration? I could not find an estimate of roughness fractal dimensions for bare silicate rock surfaces or high elevation terranes in the literature, but I suspect 2.15 is likely a maximum as a result of wind and water erosion under abiotic conditions. For example, with a value of 2.1, the computed BEW value is 182, a likely maximum (see Table 10 below).

In conclusion, I suspect that the lab artifact effect in raising dissolution rates of silicate minerals should be taken more seriously and new experiments with unpulverized samples should be considered.

**Table 10. Can the present global BEW be inferred?**

| Line | Fractal dimension |
|---|---|
| a | 2.15 (Abiotic land surface ?) |
| b | 2.3 (Earth's land surface) |

At the Watershed Level:
The difference in denudation rates between lines a and b gives a BEW = $10^{1.61}$ = 41, between b and a line with fractal dimension equal to 2.1 give a BEW = $10^{2.26}$ = 182

Note that line b passes through data point 9 which corresponds to a hot humid tropical climate (Java). Therefore these BEW estimates are likely upper limits for the present global weathering regime.

**BEW progressively increases as scale increases**

BEW is higher on average for watersheds than for clast weathering because of these factors that kick in at the greater scale:

1) Higher biotic productivity, soil stabilization is generally higher in watersheds, as well as smaller average grain size of CaMg silicates, thereby generating more surface area/volume and greater opportunity for microbial/organic acid interaction. Weathering rinds on clasts are likely closer to 2D systems, analogous to lichen-colonized bedrock exposures.

2) Because of higher biotic productivity in soils underlying forest and grassland ecosystems than more poorly vegetated alluvial fans or glacial outwash sites there should be more water retention and higher $pCO_2$ levels in watershed soils, as well as higher turnover rates for watershed soils, mixing the organics with the silicates because of greater subsurface biotic productivity.

3) The biotic sink effect should be greater for watershed soils, particularly those forming on

slopes, hence a greater flux out of the system of dissolved CaMg and bicarbonates as well as organics. Note: Only the lab data is close to abiotic, with a potential artifact effect from grinding elevating measured dissolution rates (and BET area?).

Biology has increased the surface roughness of the land by its invasion of the crust and coupling with the atmosphere and climate equivalent to the 5$^{th}$ hypothesis "Biology shapes the topography of the Critical Zone" (Brantley et al., 2011).

**Conclusion: What is magnitude of present BEW?**

The likely cumulative BEW at present is on the order of 10 to 100, with plant ecosystems adding roughly an order of magnitude on the previous BEW of microbial/lichen land colonized Earth, already with oxidative weathering and ice wedging as factors.

**Some challenges for further research**

1) Global mapping of silicate $CO_2$ consumption rates revealing the relative importance of each weathering regime with respect to the global sink flux; where are the hot spots, e.g., tropical volcanic islands (Rad et al., 2013; Hartmann et al., 2014; but see Rivé et al., 2013 for evidence of a direct magmatic carbon dioxide input to the carbon sink)?

2) Determining the cumulative BEW of early land procaryotes + rise of atmospheric oxygen + ice wedging; potential experimental approach with mesocosms. Are synergistic interactions important?

3) What is the role of fluctuating redox conditions in determining silicate $CO_2$ consumption rates ? (Van Cappellen, 2013)

4) Quantifying the role of BEW in Maher and Chamberlain (2014)'s weathering model:

"The solute production model does not capture all possible biogeochemical and climatic feedbacks that might operate"

5) How does the silicate weathering and reduced organic carbon sinks in the long-term carbon cycle vary with mean global temperature? Recent papers point to a weaker temperature dependence than Berner's BLAG model (see further discussion in Appendix).

6) Considering the biogeochemical cycle of phosphorus with respect to chemical weathering has promise in better understanding the biotic role in weathering (Buendía et al., 2014; Porada et al., 2014; Hartmann et al., 2014).

**Appendix**

**Revisiting the sinks with respect to the long-term carbon cycle**

**Silicate weathering sink**

I submit that the BLAG model did not take into full account the impact of global mean temperature (Tm) variation on physical erosion rates. Data compilations of chemical weathering rates of granite and basalt as functions of watershed temperature show considerable scatter (White et al., 1999; Dupre et al., 2003; Navarre-Sitchler and Brantley, 2007)). Removing the low temperature watersheds (with water frozen plausibly frozen for a significant fraction of time) flattens the dependence lines. West (2012) and Maher and Chamberlain (2014) made the important point that low erosion rates reduced chemical weathering flux dependence on temperature/runoff, strikingly demonstrated by low fluxes in flat thick soil laterite tropical watersheds. Herman et al. (2012) shows that global data sets strongly imply cooling climates lead to increased erosion rates, mainly associated with glaciated mountain ranges. This relationship implies that as Tm increases, the chemical weathering flux will be less than expected from the temperature/runoff dependence on the present Earth surface as well as from simple experimental kinetics because of decreasing erosion rates. Therefore, using a best fit line to determine chemical weathering activation energies with present watershed data should not automatically be extrapolated to higher Tm global climates. Further, abiotic laboratory dissolution experiments will lead to misleading conclusions regarding weathering flux dependence on temperature with respect to natural weathering regimes with strong biotic influences, especially the biotic enhancement of weathering. West (2012) argues that "At low erosion rates, as temperature and runoff exert progressively less influence on weathering rates [relevant to the long-term carbon cycle], the direct weathering feedback becomes weak to nonexistent." (p. 814). As Tm increases, biotic stabilization of soils should persist, though weaken at higher temperatures as the dominant biota shifts from forests/grasslands to microbial cover (so-called cryptogamic). This influence combined with disappearing glacial erosion should weaken the weathering flux dependence on temperature.

**Reduced organic carbon sink**

Galy et al. (2015) argue that physical erosion rather than terrestrial primary productivity is driving the flux of biospheric organic carbon burial in marine sediments. This conclusion combined with the climatic temperature dependence with respect to erosion rates found by Herman et al. (2012) points to a weaker reduced organic carbon sink as a function of $T_m$ than commonly assumed.